\documentstyle[aps]{revtex}
\draft

\begin{document}
\author{Wei-Min Sun, Xiang-Song Chen and Fan Wang} 
\address{Department of Physics and Center for Theoretical Physics, 
	Nanjing University, Nanjing 210093, China
        }
\title{Some Problems Concerning Interchange of Order of Integration in Functional Integral Formalism of U(1) Gauge Field Theories }
\maketitle
 
\begin{abstract}
We show that in the functional integral formalism of U(1) gauge field theory some formal manipulation such as interchange of order of integration can yield erroneous results. The example studied is analysed by Fubini theorem.

\end{abstract}

Functional integral has now become an indispensible tool in the study of field theories, especially gauge field theories. When using the formalism of functional integral we usually assume some formal manipulations to be justified, for example, interchange of order of integration. From the general theory of integration \cite{Rudin} we know that this is not always correct. Usually we do not have to worry about these matters. In this paper we will show that in some circumstances naive interchange of order of integration in the manipulation of functional integral does lead to erroneous results. 

 In the literature \cite{Kogan} there exists a kind of averaging procedure for constructing gauge-invariant quantities out of gauge-variant ones.  
This averaging procedure can be described as follows(to be specific we restrict ourself to pure U(1) gauge field theories). For an arbitrary operator $O[A]$, gauge invariant or not, we can always 
define an operator $G_O[A]=\int D\omega O[A^{\omega}]$ 
where $A$ denotes gauge fields, $A^{\omega}$ denotes the 
result of $A$ after a gauge transformation $\omega$ and $\int D\omega$ 
stands for functional integration over the gauge group(here we choose the normalization to be $\int D\omega=1$). Using the property of the Haar measure it can be easily proved that $G_O[A]$ is gauge-invariant. In \cite{SCW} we proved that such a method is inapplicable for $O[A]$ which is a {\em gauge-variant} polynomial in the gauge field $A_{\mu}(x)$ because the relevant functional integral in $\omega$ is divergent. 

When this averaging procedure is applicable it is obvoius that the correlation function
\begin{equation}
 \langle G_O[A]\rangle=\int D\omega \langle O[A^{\omega}]\rangle 
\end{equation}
 is gauge independent because the correlation function of any gauge-invariant operator is always gauge independent \cite{Weinberg}. Sometimes the operator $G_O[A]=\int D\omega O[A^{\omega}]$
itself does not exist  but the functional integral in the right hand side of eq.(1) {\em is} convergent. Then is the quantity $\int D\omega \langle O[A^{\omega}]\rangle$ gauge independent?

 Naively one may expect this quantity to be gauge independent because we have averaged over the whole gauge group. In the following we will use the formalism of functional integral to give a formal proof to "show" that the quantity studied is gauge independent, irrespective of whether the operator $G_O[A]$ itself exists or not. We will only investigate those $O[A]$ which can be expressed as polynomials in the gauge field $A_{\mu}(x)$. In our case of pure U(1) gauge field theory there are no problems such as Gribov ambiguity and the difficulty of rigorously defining functional measure for interacting quantum field theories. We will work in Euclidean space throughout.
 
 Recalling the Faddeev-Popov trick in quantizing a general gauge field theory we introduce a functional $\Delta_F[A]$ by the relation $\Delta_F[A]\int D\omega \delta(F[A^{\omega}])=1$, where $F[A]$ is a suitable(linear) gauge fixing functional(here we choose the normalization to be $\int D\omega=1$; this is different from the usual one but only affect an irrelevant normalization constant.), and get
\begin{eqnarray}
&&\int D\omega \int DA O[A^{\omega}]\Delta_F[A]\delta(F[A])e^{-S[A]} \nonumber \\
&=& \int D\omega \int DA O[A]\Delta_F[A]\delta(F[A^{\omega^{-1}}])e^{-S[A]} \nonumber \\
&=& \int DA \int D\omega O[A]\Delta_F[A]\delta(F[A^{\omega^{-1}}])e^{-S[A]} \nonumber \\
&=& \int DA \int D\omega O[A]\Delta_F[A]\delta(F[A^{\omega}])e^{-S[A]} \nonumber \\
&=& \int DA O[A]e^{-S[A]}
\end{eqnarray}
where in the second line we have changed the variable from $A$ to $A^{\omega^{-1}}$ and used the gauge invariance of $DA$, $\Delta_F[A]$ and the action $S[A]$, in the third line we have changed the order of integration and in the fourth line we have used the property $\int D\omega f[\omega]=\int D\omega f[\omega^{-1}]$. 
 
Now in the above equation we change $F[A]$ into $F[A]-f$, where $f$ is an arbitary function, and obtain
\begin{equation}
\int D\omega\int DA O[A^{\omega}]\Delta_F[A,f]\delta(F[A]-f)e^{-S[A]}=\int DA O[A]e^{-S[A]}
\end{equation}
Noting that $\Delta_F[A,f]\delta(F[A]-f)=\tilde{\Delta}_F[A]\delta(F[A]-f)$, where $\tilde{\Delta}_F[A]=\det\frac{\delta F[A^{\omega}]}{\delta\omega}|_{\omega=1}$ is the familiar Faddeev-Popov determinant, multiplying both sides with a suitable weight functional $G[f]\sim e^{-\frac{1}{2\lambda}\int d^4x [f(x)]^2}$(normalized to $\int Df G[f]=1$) and integrating with respect to $f$, we obtain
\begin{equation}
\int D\omega \int DA O[A^{\omega}]\tilde{\Delta}_F[A]G[F[A]]e^{-S[A]} =\int DA O[A]e^{-S[A]}   
\end{equation}
Setting $O[A]=1$ in the above equation we get
\begin{equation}
\int D\omega \int DA \tilde{\Delta}_F[A]G[F[A]]e^{-S[A]}=\int DA e^{-S[A]}
\end{equation}
Recalling that 
\begin{equation}
\langle O[A^{\omega}]\rangle=\frac{\int DA O[A^{\omega}]\tilde{\Delta}_F[A]G[F[A]]e^{-S[A]}}{\int DA\tilde{\Delta}_F[A]G[F[A]]e^{-S[A]}}
\end {equation}
we have
\begin{equation}
\int D\omega \langle O[A^{\omega}]\rangle=\frac{\int DA O[A]e^{-S[A]}}{\int DA e^{-S[A]}}
\end{equation}
That is to say, the quantity $\int D\omega \langle O[A^{\omega}]\rangle$ is gauge independent.

Is this conclusion right? Let us see an explicit example. We take $O[A]=A_{\mu}(x)F_{\nu\rho}(y)$. Note that according to the general result in \cite{SCW} $G_O[A]$ does not exist. It can be easily seen that $\langle O[A^{\omega}]\rangle=\langle O[A]\rangle$ because $\langle F_{\nu\rho}(y)\rangle$ vanishes. But the value of $\langle A_{\mu}(x)F_{\nu\rho}(y)\rangle$ differs in covariant and axial gauges and so the above conclusion is wrong!   
     
Then where does the above formal proof go wrong? The point is that in the third line of eq.(2) we have illegitimately interchanged the order of integration. When the operator $G_O[A]$ exists, that is, the functional integral $\int D\omega O[A^{\omega}]$ converges, there is no problem. But when $\int D\omega O[A^{\omega}]$ diverges problems do arise. 

If the functional integral $\int D\omega O[A^{\omega}]$ diverges, the iterated integral $\int DA \int D\omega O[A^{\omega}]\Delta_F[A]\delta(F[A])e^{-S[A]}$ also does not exist.  Then from Fubini theorem \cite{Rudin} we know that the double integral $\int\int D\omega DA O[A^{\omega}]\Delta_F[A]\delta(F[A])e^{-S[A]}$ does not exist. Now we do a change of variable: $A^{\omega}=\tilde{A}, \omega=\tilde{\omega}$. The Jacobian matrix $J$ of this change of variable can be written as
\[
\left(
\begin{array}{cc}
\frac{\partial\tilde{A}}{\partial A} & \frac{\partial\tilde{A}}{\partial \omega} \\
0 & 1
\end{array}
\right)
\]
and $\det J=\det\frac{\partial\tilde{A}}{\partial A}=1$. Under this change of variable the double integral changes into $\int\int D\omega DA O[A]\Delta_F[A]\delta(F[A^{\omega^{-1}}])e^{-S[A]}$(we have changed the dummy variables $(\tilde{A}, \tilde{\omega})$ back into $(A, \omega)$). Obviously this double integral does not exist because the former one does not. In this case the two iterated integrals $\int D\omega \int DA O[A]\Delta_F[A]\delta(F[A^{\omega^{-1}}])e^{-S[A]}$ and $\int DA \int D\omega O[A]\Delta_F[A]\delta(F[A^{\omega^{-1}}])e^{-S[A]}$ are not necessarily equal, even both of them exist.      

The above analysis tells us that when the functional integral $\int D\omega O[A^{\omega}]$ does not exist the formal proof of the gauge independence of the quantity $\int D\omega \langle O[A^{\omega}]\rangle$ does not apply. It was proved in \cite{SCW} that when $O[A]$ is a gauge-variant polynomial in $A_{\mu}(x)$ the operator $G_O[A]$ does not exist. Now let us assume $O[A]$ is gauge-variant and directly analyse whether $\int D\omega \langle O[A^{\omega}]\rangle$ exists, and when exists, whether it is gauge independent. 

To investigate this problem recall that in \cite{SCW} we have proved that the functional integral $\int D\omega \partial_{\mu}\theta(x)$ does not exist. Using similar reasoning it can be shown that the functional integral $\int D\omega \int dx K^{\mu}(x)\partial_{\mu}\theta(x)$ does not exist for a general $K^{\mu}(x)$. Now we will show that the functional integral $\int D\omega P[\partial_{\mu}\theta(x)]$ also does not exist for any polynomial $P[\partial_{\mu}\theta(x)]$. The proof is by mathematical induction. We already know that this statement is true for any first order polynomial in $\partial_{\mu}\theta(x)$. Assume this statement holds for any $n$-th order polynomial $P_n[\partial_{\mu}\theta(x)]$, let us see an $(n+1)$-th order polynomial $P_{n+1}[\partial_{\mu}\theta(x)]$. If the functional integral $\int D\omega P_{n+1}[\partial_{\mu}\theta(x)]$ exist, it should be equal to the functional integral $\int D\omega P_{n+1}[\partial_{\mu}\theta(x)+\partial_{\mu}\theta_0(!
x)]$,where $\theta_0(x)$ is an arbitary function. From these we find the functional integral $\int D\omega \{P_{n+1}[\partial_{\mu}\theta(x)+\partial_{\mu}\theta_0(x)]-P_{n+1}[\partial_{\mu}\theta(x)]\}$ exists and equals zero. But note that the integrand of this functional integral is an $n$-th order polynomial in $\partial_{\mu}\theta(x)$, hence a contradiction with our assumption. That is to say, the functional integral $\int D\omega P_{n+1}[\partial_{\mu}\theta(x)]$ also does not exist. From the principle of mathematical induction our statement is proved.

Now let us turn to the operator $\int D\omega O[A^{\omega}]$ and the quantity $\int D\omega\langle O[A^{\omega}]\rangle$. When the operator $O[A]$ is a gauge-variant polynomial in $A_{\mu}(x)$ the operator $O[A^{\omega}]$ and the correlation function $\langle O[A^{\omega}]\rangle$ should be a polynomial in $\partial_{\mu}\theta(x)$. According to the above proved result the functional integral $\int D\omega O[A^{\omega}]$ does not exist(note that this provides another proof of the final conclusion in \cite{SCW}). The functional integral $\int D\omega \langle O[A^{\omega}]\rangle$ also does not exist in general. But there is a special case: $\langle O[A^{\omega}]\rangle$ is a constant independent of $\partial_{\mu}\theta(x)$, i.e.,$\langle O[A^{\omega}]\rangle=\langle O[A]\rangle$(our explicit example $A_{\mu}(x)F_{\nu\rho}(y)$ is such a case).
In other words the gauge variation of $O[A]$ vanishes after taking the vacuum expectation value. In this case the quantity $\int D\omega\langle O[A^{\omega}]\rangle$ does exist; but unfortunately it is generally gauge dependent because $O[A]$ is a gauge-variant operator. Returning back to the problem of whether we can interchange the order of integration in our functional proof we conclude that this formal manipulation is erroneous in this case.

In summary we showed that when $O[A]$ is a {\em gauge-variant} polynomial in the gauge field $A_{\mu}(x)$ the quantity $\int D\omega \langle O[A^{\omega}]\rangle$ is gauge dependent in general(if it exists) even though we have averaged over the gauge group. The formal functional proof for the gauge independence of this quantity does not apply because one have illegitimately changed the order of integration when doing the functional integral.
         
This work is supported in part by the NSF(19675018), SED and SSTD of China.


\begin{thebibliography}{99}
\bibitem{Rudin} See, for inatance, W.Rudin {\it Real and Complex Analysis, 3rd edition, Ch8} (McGraw-Hill, 1987)          
\bibitem{Kogan} I.I.Kogan and A.Kovner, Phys. Rev. {\bf D51} (1995) 1948. 
\bibitem{SCW} W.M.Sun, X.S.Chen and F.Wang, Phys. Lett. {\bf B483} (2000) 299, hep-th/0005060 
\bibitem{Weinberg} See, for instance, S. Weinberg {\it The Quantum Theory of Fields, vol 2, Ch 15} (Cambridge University Press, 1996) 
\end{thebibliography}
\end{document}